\documentclass{fac}
\usepackage{graphicx}
\usepackage{amssymb}
\usepackage{amsfonts}
\usepackage{amsmath}
\usepackage{dsfont}
\ifprodtf \else \usepackage{latexsym}\fi

\newcommand\black{\ensuremath{\blacktriangleright}}
\newcommand\white{\ensuremath{\vartriangleright}}

\newif\ifamsfontsloaded
\ifprodtf
  \newcommand\whbl{\white\kern-.1em--\kern-.1em\black}
  \newcommand\blwh{\black\kern-.1em--\kern-.1em\white}
  \newcommand\blbl{\black\kern-.1em--\kern-.1em\black}
  \newcommand\whwh{\white\kern-.1em--\kern-.1em\white}
  \amsfontsloadedtrue
\else
  \checkfont{msam10}
  \iffontfound
    \IfFileExists{amssymb.sty}
      {\usepackage{amssymb}\amsfontsloadedtrue
       \newcommand\whbl{\white\kern-.125em--\kern-.125em\black}%
       \newcommand\blwh{\black\kern-.125em--\kern-.125em\white}%
       \newcommand\blbl{\black\kern-.125em--\kern-.125em\black}%
       \newcommand\whwh{\white\kern-.125em--\kern-.125em\white}}
      {}
  \fi
\fi

\newtheorem{theorem}{Theorem}[section]
\newtheorem{definition}{Definition}[section]

\title[A Process Algebra for Games]
      {A Process Algebra for Games}

\author[Yong Wang]
    {Yong Wang\\
     College of Computer Science and Technology,\\
     Faculty of Information Technology,
     Beijing University of Technology, Beijing, China\\
     }

\correspond{Yong Wang, Pingleyuan 100, Chaoyang District, Beijing, China.
            e-mail: wangy@bjut.edu.cn}

\pubyear{2013}
\pagerange{\pageref{firstpage}--\pageref{lastpage}}

\begin{document}
\label{firstpage}

\makecorrespond

\maketitle

\begin{abstract}
Using formal tools in computer science to describe games is an interesting problem. We give games, exactly two person games, an axiomatic foundation based on the process algebra ACP (Algebra of Communicating Process). A fresh operator called opponent's alternative composition operator (OA) is introduced into ACP to describe game trees and game strategies, called GameACP. And its sound and complete axiomatic system is naturally established. To model the outcomes of games (the co-action of the player and the opponent), correspondingly in GameACP, the execution of GameACP processes, another operator called playing operator (PO) is extended into GameACP. We also establish a sound and complete axiomatic system for PO. To overcome the new occurred non-determinacy introduced by GameACP, we extend truly concurrent process algebra APTC for games called GameAPTC. Finally, we give the correctness theorem between the outcomes of games and the deductions of GameACP and GameAPTC processes.
\end{abstract}


\section{Introduction}

Game theory \cite{Games} is a great theoretical outcome of the 20th century and is used widely in interpreting social and economic phenomena. Because of the universality of games, games are also widely used in science and engineering. On one side, game theory has been applied in many domains successfully. On the other side, using different tools to interpret game theory is also an interesting direction.

In computer science, there are various kind of tools to capture the computation concept concerned with the nature of computability. There is no doubt that process algebra \cite{PA} is one of the most influential tools. Milner's CCS (Calculus of Communicating Systems) \cite{CCS}, Hoare's CSP (Communicating sequential processes) \cite{CSP} and ACP (Algebra of Communicating Process) \cite{ACP} are three dominant forms of process algebra, and there are also several kinds of process calculi. Such process algebras often have a formal deductive system based on equational logic and a formal semantics model based on labeled transition systems, can be suitable to reason about the behaviors of parallel and distributed systems.

The combination of games and computer science is a fascinating direction, and it gains great successes, such as the so-called game semantics \cite{PCF3}. Since there exist lots of game phenomena in parallel and distributed systems, especially interactions between a system and its environment, interactions among system components, and interactions among system components and outside autonomous software agents, the introduction of games into traditional computation tools, such as the above mentioned process algebra, is attractive and valuable. The  computation tools extended to support games can be used to reason about the behaviors of systems in a new viewpoint.

Using these computation tools to give game theory an interpretation is an interesting problem \cite{GameProcess}. This direction has subtle difference with introducing games and ideas of games into the computation tools. It not only can make these tools having an additional ability to using games in computation, but also gives game theory a new interpretation which will help the human to capture the nature of games and also the development of game theory.

Although in some process algebras, such as CSP, there are an internal choice $\sqcap$ and an external choice $\square$, process algebra ACP does not distinguish the internal choice and the external choice. In this paper, we introduce external choice into process algebra ACP in a game theory flavor. we give games an axiomatic foundation called GameACP based on process algebra ACP \cite{ACP}. Because of ACP's clear semantic model based on bisimulation or rooted branching bisimulation \cite{SOS} and well designed axiomatic system, GameACP inherits ACP's advantages in an elegant and convenient way. This is the first step to use computation tools to interpret games in an axiomatic fashion as far as we know.

This paper is organized as follows. In Section 2, we analyze the related works. Application scenarios called SubmittingOrder, Transacting and Purchasing are illustrated in Section 3. In Section 4, we briefly introduce some preliminaries, including process algebra ACP and also games. In Section 5, the extension of BPA (Basic Process Algebra) for games is done, which is called GameBPA, including opponent's alternative composition operator and another new operator called playing operator of GameACP processes, and their transition rules and the properties of the extension, and we design the axioms of opponent's alternative composition and playing operator, including proving the soundness and completeness of the axiomatic system. In Section 6, we do another extension based on ACP, which is called GameACP. In Section 7, we do another extension based on APTC, which is called GameAPTC. We give the correctness theorem in Section 8. In Section 9, we show the support for multi-person games. Finally, conclusions are drawn in Section 10.

\section{Related Works}\label{RelatedWorks}

As mentioned above, the combination of computation tools and game semantics includes two aspects: one is introducing games or idea of games into these computation languages or tools to give them a new viewpoint, and the other is using these computation tools to interpret games. The first one has plenty of works and gained great successes, but the second one has a few works \cite{GameCTR} \cite{GameProcess} as we known. We introduce the two existing works in the following.

It is no doubt that the so-called game semantics gained the most great successes in introducing games into computer science. Game semantics models computations as playing of some kind of games, especially two person games. In the two person game, the Player (P) represents the system under consideration and the Opponent (O) represents the environment in which the system is located. In game semantics, the behaviors of the system (acts as P) and the environment (acts as O) are explicitly distinguished. So the interactions between the system and the environment can be captured as game plays between the opponent and the player, and successful interactions can be captured by the game strategy.

\begin{figure}
  \centering
  %\vspace{5cm}
  \includegraphics{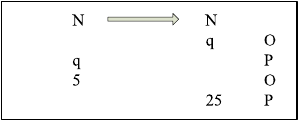}
  \caption{Game Semantics for the Function $f(x)=3x+10$ where $x \in \mathds{N}$.}
  \label{F(x)}
\end{figure}

For example, the function $f(x)=3x+10$ where $x \in \mathds{N}$ can be deemed as the games played in Fig. \ref{F(x)}. Firstly, the opponent (the environment) moves to ask the value of $f(x)$, then the player (the function) moves to ask the value of $x$, and then the opponent moves to answer that the value of $x$ is 5, the player moves to answer that the value of $f(x)$ is 25 finally.

Game semantics has gained great successes in modeling computations, such as an initial success of modeling the functional programming language PCF (Programming Computable Functions) \cite{PCF} \cite{PCF2} \cite{PCF3}, multiplicative linear logic \cite{MIL}, idealized Algol \cite{Algol}, general reference \cite{GR}, etc. To model concurrency in computer science with game semantics, a new kind of game semantics called asynchronous game \cite{AsynGame1} \cite{AsynGame2} \cite{AsynGame3} \cite{AsynGame4} \cite{AsynGame5} is established and a bridge between the asynchronous game and traditional game semantics is founded. Moreover, asynchronous games perfectly model propositional linear logic and get a full completeness result. Another kind of game semantics to describe concurrency is concurrent game \cite{ConGame1} \cite{ConGame2}, and a work to bridge asynchronous game and concurrent game is introduced in \cite{AsynAndConGame}.

Algorithmic game semantics \cite{Algorithm} is the premise of implementation of game semantics for further automatic reasoning machine based on some specific game semantics model. And game semantics can be used to establish the so-called interaction semantics \cite{Interaction} among autonomous agents, and can be used to model and verify compositional software \cite{COMPOSITION1} \cite{COMPOSITION2}.

Game semantics utilizes such dialogue games to model interactions between the system under consideration and the environment, and pays more attention to the playing process of the two players. And it develops some key concepts which have correspondents to traditional computation concepts, such as innocence to context independence and bracketing to well-structured property. Different to game semantics, there are also several works to use computation tools to model games of two agents.

Game-CTR \cite{GameCTR} introduces games into CTR (Concurrent Transaction Logic) to model and reason about runtime properties of workflows that are composed of non-cooperative services -- such as Web Services. Game-CTR includes a model and proof theory which can be used to specify executions under some temporal and causality constraints, and also a game solver algorithm to convert such constraints into other equivalent Game-CTR formulas to be executed more efficiently. Chatzikokolakis et al\cite{GameProcess} develop a game semantics for a certain kind of process calculus with two interacting agents. Games and strategies on this process calculus are defined, and strategies of the two agents determine the execution of the process. And also, a certain class of strategies correspond to the so-called syntactic schedulers of Chatzikokolakis and Palamidessi. In these works, the games used are not dialogue games, and there are no interactions such as questions and answers and also no wining concept.

More like Game-CTR \cite{GameCTR} and Chatzikokolakis's work \cite{GameProcess}, we introduce games into ACP, or we use ACP to give games an interpretation. Unlike \cite{GameCTR} and \cite{GameProcess}, our work GameACP and GameAPTC is an attempt to do axiomatizations with extensions of process algebra ACP and APTC for games. It has the following characteristics:

\begin{enumerate}
  \item We introduce the external choice into process algebra ACP and APTC in a game theory flavor. As a result of axiomatization, GameACP and GameAPTC have not only an equational logic, but also a bisimulation semantics.
  \item The conclusions of GameACP and GameAPTC are without any assumption or restriction, such as epistemic restrictions on strategies in \cite{GameProcess}.
  \item Though the discussions of GameACP and GameAPTC are aimed at two person games, GameACP and GameAPTC can be naturally used in multi-person games.
  \item GameACP and GameAPTC provide new viewpoints to model interactions between one autonomous agent and other autonomous agents, and can be used to reason about the behaviors of parallel and distributed systems with game theory supported.
\end{enumerate}

\section{Application Scenarios}\label{AS}

In this section, we will illustrate the universality of game phenomena that exist in computer systems through three different examples. Using these examples throughout this paper, we illustrate our core concepts and ideas.

\subsection{Graphical User Interface -- SubmittingOrder}

\begin{figure}
  \centering
  %\vspace{5cm}
  \includegraphics{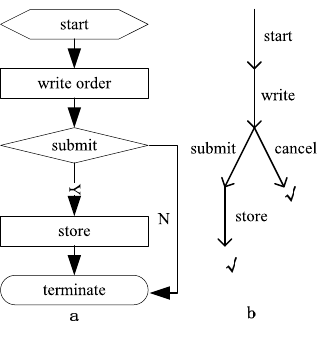}
  \caption{Submitting An Order Example.}
  \label{SubmittingOrder}
\end{figure}

Graphical user interface is the most popular human-machine interface now. Fig. \ref{SubmittingOrder}-a illustrates the flow of submitting an order for a user through a graphical interface. The flow is as follows.

\begin{enumerate}
  \item The interface program starts.
  \item The user writes an order via the interface.
  \item When the order is completed, the user can decide to submit the order or cancel the order.
  \item If the order is submitted, then the order is stored and the program terminates.
  \item If the order is canceled, then the program terminates.
\end{enumerate}

In this SubmittingOrder example, the selection of submitting or canceling the order is done by the user, but not the program according to its inner states. This situation is suitable to be captured by use of a game between the user and the interface program.

\subsection{Transaction Processing -- Transacting}

\begin{figure}
  \centering
  %\vspace{5cm}
  \includegraphics{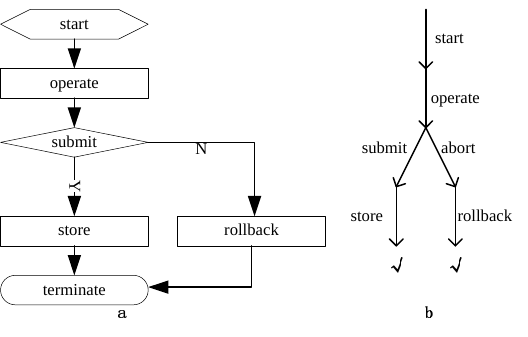}
  \caption{Database Transaction Processing Example.}
  \label{Transaction}
\end{figure}

Transaction processing is the core mechanism of database and business processing. Traditional transaction has ACID properties and is illustrated in Fig. \ref{Transaction}-a. The flow of traditional database transaction is following.

\begin{enumerate}
  \item The transaction is started.
  \item Operations on the data are done by a user.
  \item The user can decide to submit the transaction or abort the transaction.
  \item If the transaction is submitted, the data are permanently stored and the transaction terminates.
  \item If the transaction is aborted, the data are rollbacked and the transaction also terminates.
\end{enumerate}

In this Transaction example, the selection of submitting or aborting the transaction is also done by the user, but not the database or business processing system according to its inner states. This situation is also suitable to be modeled by use of a game between the user and the database or business processing system.

\subsection{Web Service Composition -- Purchasing}

Web Service is a quite new distributed object and Web Service composition created new bigger Web Services from the set of smaller existing Web Services. A composite Web Service is defined by use of a kind of Web Service composition language and is executed by interpreting the definition of the composite Web Service. WS-BPEL\cite{WSBPEL} is a kind of such language. In WS-BPEL, the atomic function units are called atomic activities and the corresponding structural activities define the control flow among these atomic activities. \textbf{Pick} activity is a kind of choice structural activity in which the decision is made by outside autonomous Web Services, and is different from the \textbf{If} activities, in which the decision is made by the composite Web Service according to its inner states.

\begin{figure}
  \centering
  %\vspace{5cm}
  \includegraphics{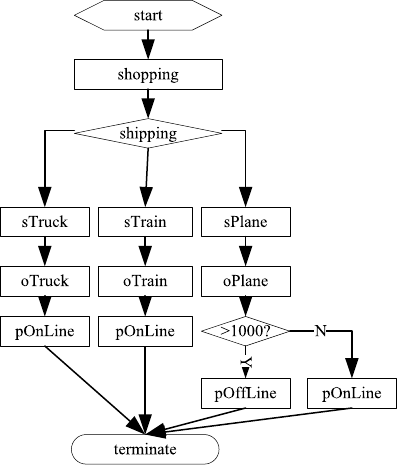}
  \caption{Flows of Purchasing Goods Example.}
  \label{Purchasing1}
\end{figure}

In Fig. \ref{Purchasing1}, a composite Web Service implements the following flow of purchasing goods and can be used by a user through a user agent Web Service.

\begin{enumerate}
  \item The composite Web Service is started by a user through a user agent Web Service.
  \item The user shops for goods.
  \item After the shopping is finished, the user can select the shipping way: by truck, by train or by plane.
  \item If the truck way is selected, then the user should order a truck and pay online for the fees.
  \item If the train way is selected, then user should order a train and pay online for the fees.
  \item If the plane way is selected, then the user should order a plane, if the money amount is greater than 1000 dollars, he/she should pay offline for the fees; and if not, he/she should pay online.
\end{enumerate}

\begin{figure}
  \centering
  %\vspace{5cm}
  \includegraphics{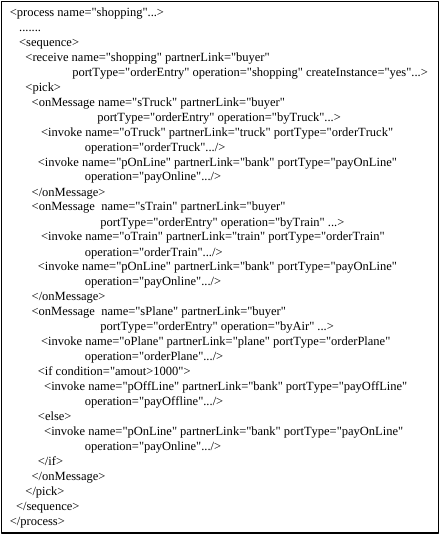}
  \caption{WS-BPEL Skeleton of Purchasing Goods Example.}
  \label{Purchasing2}
\end{figure}

The WS-BPEL skeleton of the Purchasing composite Web Service is shown in Fig. \ref{Purchasing2}. Note that the first choice is modeled by use of a \textbf{Pick} activity and the second choice is modeled by use of an \textbf{If} activity.

In this Purchasing composite Web Service, the selection of shipping ways is also done by the user through a user agent Web Service, and not the composite Web Service according to its inner states. This situation is also suitable to be modeled by use of a game between the user (or the user agent Web Service) and the composite Web Service.

\section{Preliminaries}

In this section, we introduce some preliminaries, including process algebra ACP and games, on equational logic and structured operational semantics, please refer to \cite{ACP}.

In the following, the variables $x,x',y,y',z,z'$ range over the collection of process terms, the variables $\upsilon,\omega$ range over the set $A$ of atomic actions, $a,b,c\in A$, $s,s',t,t'$ are closed items, $\tau$ is the special constant silent step, $\delta$ is the special constant deadlock, and the predicate $\xrightarrow{a}\surd$ represents successful termination after execution of the action $a$.

\subsection{Process Algebra -- ACP}\label{ACPP}

ACP\cite{ACP} is a kind of process algebra which focuses on the specification and manipulation of process terms by use of a collection of operator symbols. In ACP, there are several kind of operator symbols, such as basic operators to build finite processes (called BPA), communication operators to express concurrency (called PAP), deadlock constants and encapsulation enable us to force actions into communications (called ACP), linear recursion to capture infinite behaviors (called ACP with linear recursion), the special constant silent step and abstraction operator (called $ACP_{\tau}$ with guarded linear recursion) allows us to abstract away from internal computations.

Bisimulation or rooted branching bisimulation based structural operational semantics is used to formally provide each process term used the above operators and constants with a process graph. The axiomatization of ACP (according the above classification of ACP, the axiomatizations are $\mathcal{E}_{\textrm{BPA}}$, $\mathcal{E}_{\textrm{PAP}}$, $\mathcal{E}_{\textrm{ACP}}$, $\mathcal{E}_{\textrm{ACP}}$ + RDP (Recursive Definition Principle) + RSP (Recursive Specification Principle), $\mathcal{E}_{\textrm{ACP}_\tau}$ + RDP + RSP + CFAR (Cluster Fair Abstraction Rule) respectively) imposes an equation logic on process terms, so two process terms can be equated if and only if their process graphs are equivalent under the semantic model.

ACP can be used to formally reason about the behaviors, such as processes executed sequentially and concurrently by use of its basic operator, communication mechanism, and recursion, desired external behaviors by its abstraction mechanism, and so on.

ACP can be extended with fresh operators to express more properties of the specification for system behaviors. These extensions are required both the equational logic and the structural operational semantics to be extended. Then the extension can be done based on ACP, such as its concurrency, recursion, abstraction, etc.

\subsubsection{SubmittingOrder Described by ACP}

The process graph of the SubmittingOrder example is illustrated in Fig. \ref{SubmittingOrder}-b. Since ACP does not distinguish the choice decision made by outside agent or inner states, the process of the SubmittingOrder example can be expressed by the following process term in ACP.

$start\cdot write\cdot (submit\cdot store + cancel)$.

\subsubsection{Transaction Described by ACP}

The process graph of the Transaction example is illustrated in Fig. \ref{Transaction}-b. The process of the Transaction example can be expressed by the following process term in ACP.

$start\cdot operate\cdot (submit\cdot store + abort\cdot rollback)$.

\subsubsection{Purchasing Described by ACP}

\begin{figure}
  \centering
  %\vspace{5cm}
  \includegraphics{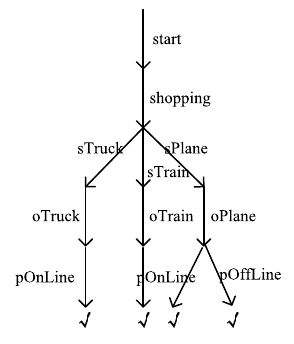}
  \caption{Process Graph of Purchasing Goods Example.}
  \label{Purchasing3}
\end{figure}

The process graph of the Purchasing composite Web Service is illustrated in Fig. \ref{Purchasing3}. The process of the Purchasing composite Web Service can be expressed by the following process term in ACP.

$start\cdot shopping\cdot (sTruck\cdot oTruck\cdot pOnLine + sTrain\cdot oTrain\cdot pOnLine + sPlane \cdot oPlane \cdot (pOnLine + pOffLine))$.

\subsection{Games}\label{Games}

In the above application scenarios, one agent interacts with other autonomous agents or human beings. In the agent's viewpoint, some branch decisions are made by outside agents or human beings, but not the inner states. In this situation, a two person game is suitable to model the interaction. In the game, the agent is modeled as the Player (denoted as P) and the other agent or the human being is modeled as the opponent (denoted as O).

\begin{figure}
  \centering
  %\vspace{5cm}
  \includegraphics{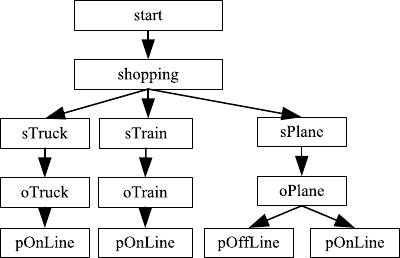}
  \caption{Game Tree of Purchasing Goods Example.}
  \label{PurchasingGameTree}
\end{figure}

Corresponding to a process graph, there exists a game tree, for example, the game tree corresponding to process graph in Fig. \ref{Purchasing3} is illustrated in Fig. \ref{PurchasingGameTree}.

We define move and strategy as follows.

\begin{definition}[Move]
Every execution of an action $a$ in the process graph causes a move $a$ in the corresponding game tree.
\end{definition}

And we do not distinguish the action $a$ and the move $a$.

\begin{definition}[P strategy]
A strategy $\lambda_P$ of P in a game tree is a subtree defined as follows:

\begin{enumerate}
  \item the empty move $\epsilon \in \lambda_P$;
  \item if the move $m \in \lambda_P$ is a P move, then exactly one child move $m'$ of $m$ and $m' \in \lambda_P$;
  \item if the move $m \in \lambda_P$ is an O move, then all children $M'$ of $m$ are in $\lambda_P$, that is $M' \subseteq \lambda_P$.
\end{enumerate}
\end{definition}

Since P and O are relative, the strategy $\lambda_O$ of O can be defined similarly.

\begin{definition}[O strategy]
A strategy $\lambda_O$ of O in a game tree is a subtree define as follows:

\begin{enumerate}
  \item the empty move $\epsilon \in \lambda_O$;
  \item if the move $m \in \lambda_O$ is a O move, then exactly one child move $m'$ of $m$ and $m' \in \lambda_O$;
  \item if the move $m \in \lambda_O$ is an P move, then all children $M'$ of $m$ are in $\lambda_O$, that is $M' \subseteq \lambda_O$.
\end{enumerate}
\end{definition}

In the game tree illustrated in Fig. \ref{PurchasingGameTree} of Purchasing example, there are two choice decisions. One is made by the user agent (or the user), and the other is made by the composite service. In this game, we model the composite service as P and the user agent (or the user) as O.

\begin{figure}
  \centering
  %\vspace{5cm}
  \includegraphics{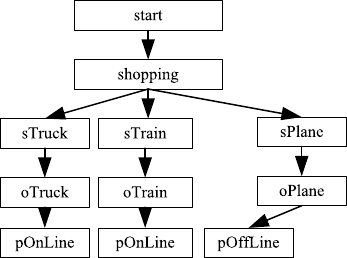}
  \caption{Player Strategy of Purchasing Game.}
  \label{PurchasingPlayer}
\end{figure}

\begin{figure}
  \centering
  %\vspace{5cm}
  \includegraphics{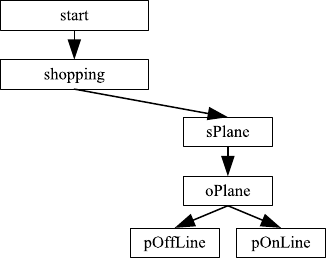}
  \caption{Opponent Strategy of Purchasing Game.}
  \label{PurchasingOpponent}
\end{figure}

A strategy of P is illustrated as Fig. \ref{PurchasingPlayer} shows. And a strategy of O is as Fig. \ref{PurchasingOpponent} illustrates.

We can see that the actual execution a game tree are acted together by the P and the O. For a P strategy $\lambda_P$ and an O strategy $\lambda_O$ of a game tree, $\lambda_P \cap \lambda_O$ has the form $\lambda_P \cap \lambda_O=\{\epsilon,m_1,m_1\cdot m_2,...,m_1\cdot...\cdot m_n\}$ according to the definition of strategy. We can get that the maximal element $m_1\cdot...\cdot m_n$ of $\lambda_P \cap \lambda_O$ exactly defines an execution of the game tree.

\begin{figure}
  \centering
  %\vspace{5cm}
  \includegraphics{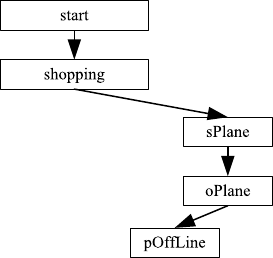}
  \caption{Execution of Purchasing Process.}
  \label{Execution}
\end{figure}

For the P strategy $\lambda_P$ illustrated in Fig. \ref{PurchasingPlayer} and O strategy $\lambda_O$ illustrated in Fig. \ref{PurchasingOpponent}, the maximal element $start\cdot shopping\cdot sPlane \cdot oPlane \cdot pOffLine$ of $\lambda_P \cap \lambda_O$ defines an execution of the process as illustrated in process graph Fig. \ref{Purchasing3}. This is shown in Fig. \ref{Execution}.

\section{Extension of BPA for Games -- GameBPA}\label{extension}
GameBPA is based on BPA. In BPA, there are two basic operators called alternative composition $+$ and sequential composition $\cdot$. We give the transition rules for BPA as follows.

$$
\frac{}{\upsilon\xrightarrow{\upsilon}\surd}
$$

$$
\frac{x\xrightarrow{\upsilon}\surd}{x + y\xrightarrow{\upsilon}\surd}
$$

$$
\frac{x\xrightarrow{\upsilon}x'}{x + y\xrightarrow{\upsilon}x'}
$$

$$
\frac{y\xrightarrow{\upsilon}\surd}{x + y\xrightarrow{\upsilon}\surd}
$$

$$
\frac{y\xrightarrow{\upsilon}y'}{x + y\xrightarrow{\upsilon}y'}
$$

$$
\frac{x\xrightarrow{\upsilon}\surd}{x \cdot y\xrightarrow{\upsilon}y}
$$

$$
\frac{x\xrightarrow{\upsilon}x'}{x \cdot y\xrightarrow{\upsilon}x'\cdot y}
$$

The axioms of BPA are in Table \ref{AxiomOfBPA}.

\begin{center}
\begin{table}
  \begin{tabular}{@{}ll@{}}
\hline No. &Axiom\\
  A1 & $x + y = y + x$ \\
  A2 & $(x + y) + z = x + (y + z)$ \\
  A3 & $x + x = x$ \\
  A4 & $(x + y)\cdot z = x\cdot z + y\cdot z$ \\
  A5 & $(x\cdot y)\cdot z = x\cdot (y\cdot z)$\\
\end{tabular}
\caption{Axioms of BPA}
\label{AxiomOfBPA}
\end{table}
\end{center}

The main results on BPA are the following ones.

\begin{theorem}
Bisimulation equivalence is a congruence with respect to BPA.
\end{theorem}

\begin{theorem}
$\mathcal{E}_{\textrm{BPA}}$  is sound for BPA modulo bisimulation equivalence.
\end{theorem}

\begin{theorem}
$\mathcal{E}_{\textrm{BPA}}$ is complete for BPA modulo bisimulation equivalence.
\end{theorem}

In order to support game theory, we need an extension of ACP. We design an operator $\ddagger$ called opponent's alternative composition operator to describe the alternative decision made by an outside autonomous agent. The choice of P moves can be captured by alternative composition operator $+$ in ACP, but there is no corresponding operator to model the choice of O moves. We extend a new operator $\ddagger$ called opponent's alternative composition operator to describe the choice of O moves.

\subsection{Scenarios Described by GameBPA}

\subsubsection{SubmittingOrder Described by GameBPA}

In SubmittingOrder example, we model the interface program as P and the user as O. Then in the P's view, the process can be expressed by the following process term in GameBPA.

$start\cdot write\cdot (submit\cdot store \ddagger cancel)$.

And the subtree corresponding to the above term is the only strategy of P.

In the O's view, the process can be expressed by the following process term in GameBPA.

$start\cdot write\cdot (submit\cdot store + cancel)$.

So the subtrees corresponding to term $start\cdot write\cdot submit\cdot store$ and $start\cdot write\cdot cancel$ are all strategies of O.

\subsubsection{Transaction Described by GameBPA}

In Transaction example, we model the database as P and the user as O. Then in the P's view, the process can be expressed by the following process term in GameBPA.

$start\cdot operate\cdot (submit\cdot store \ddagger abort\cdot rollback)$.

And the subtree corresponding to the above term is the only strategy of P.

In the O's view, the process can be expressed by the following process term in GameBPA.

$start\cdot operate\cdot (submit\cdot store + abort\cdot rollback)$.

So the subtrees corresponding to term $start\cdot operate\cdot submit\cdot store$ and $start\cdot operate\cdot abort\cdot rollback$ are all strategies of O.

\subsubsection{Purchasing Described by GameBPA}

In Purchasing example, we model the composite service as P and the user agent (or the user) as O. Then in the P's view, the process can be expressed by the following process term in GameBPA.

$start\cdot shopping\cdot (sTruck\cdot oTruck\cdot pOnLine \ddagger sTrain\cdot oTrain\cdot pOnLine \ddagger sPlane \cdot oPlane \cdot (pOnLine + pOffLine))$.

So the subtrees corresponding to term $start\cdot shopping\cdot (sTruck\cdot oTruck\cdot pOnLine \ddagger sTrain\cdot oTrain\cdot pOnLine \ddagger sPlane \cdot oPlane \cdot pOnLine)$ and $start\cdot shopping\cdot (sTruck\cdot oTruck\cdot pOnLine \ddagger sTrain\cdot oTrain\cdot pOnLine \ddagger sPlane \cdot oPlane \cdot pOffLine)$ are all strategies of P.

In the O's view, the process can be expressed by the following process term in GameBPA.

$start\cdot shopping\cdot (sTruck\cdot oTruck\cdot pOnLine + sTrain\cdot oTrain\cdot pOnLine + sPlane \cdot oPlane \cdot (pOnLine \ddagger pOffLine))$.

So the subtrees corresponding to term $start\cdot shopping\cdot sTruck\cdot oTruck\cdot pOnLine$, $start\cdot shopping\cdot sTrain\cdot oTrain\cdot pOnLine$ and $start\cdot shopping\cdot sPlane \cdot oPlane \cdot (pOnLine \ddagger pOffLine)$ are all strategies of O.

\subsection{Transition Rules of Opponent's Alternative Composition Operator}

Based on the above discussions, the transition rules of opponent's alternative composition operator $\ddagger$ are given as follows. $x,x',y,y'$ range over $A$, the variables $\upsilon$ range over the set $A$ of atomic actions. We define two set of $A_{\lambda_P}$ and $A_{\lambda_O}$ which denote the set of atomic actions in $\lambda_{P}$ and the set of atomic actions in $\lambda_{O}$.

$$
\frac{x\xrightarrow{\upsilon}\surd}{x\ddagger y\xrightarrow{\upsilon}\surd}
$$

$$
\frac{x\xrightarrow{\upsilon}x'}{x\ddagger y\xrightarrow{\upsilon}x'}
$$

$$
\frac{y\xrightarrow{\upsilon}\surd}{x\ddagger y\xrightarrow{\upsilon}\surd}
$$

$$
\frac{y\xrightarrow{\upsilon}y'}{x\ddagger y\xrightarrow{\upsilon}y'}
$$

where $\upsilon\in A_{\lambda_O}$. That is, in the view of O, the opponent's alternative composition is the same as the traditional alternative composition.

In the O's view, the first transition rule of opponent's alternative composition operator $\ddagger$ says if $t$ terminates successfully by executing an action $a$ then the process term $t \ddagger s$ will terminate successfully by executing the action $a$. The second one says if $t$ evolves into $t'$ by executing an action $a$ then the process term $t \ddagger s$ will evolve into $t'$ by executing the same action $a$. The third transition rule of opponent's alternative composition operator $\ddagger$ says if $s$ terminates successfully by executing an action $a$ then the process term $t \ddagger s$ will terminate successfully by executing the action $a$. The fourth one says if $s$ evolves into $s'$ by executing an action $a$ then the process term $t \ddagger s$ will evolve into $s'$ by executing the action $a$.

But, in P's view, the choice of the opponent's alternative composition can not be done according to its own inner states or its own knowledge. That is, the choice of the opponent's alternative composition is depending on the O in the game. In the P's view, $x \ddagger y$ is non-deterministic in nature and we call it as \textbf{a GameBPA Process}. A GameBPA process can not execute according to its own knowledge without the co-action with the O. To make a GameBPA process executable, the opponent is joined and the game plays. At this time, $x \ddagger y$ is deterministic, and behaves just the same as $x + y$.

\subsection{Properties of GameBPA}

We can get the following two properties of GameBPA.

\begin{theorem}
GameBPA is a conservative extension of BPA (see Section \ref{ACPP}).
\end{theorem}

\begin{proof}
This theorem follows from the following two facts.

\begin{enumerate}
  \item The transition rules of BPA (see Section \ref{ACPP}) are all source-dependent.
  \item The sources of the four transition rules for the opponent's alternative operator all contain an occurrence of $\ddagger$.
\end{enumerate}

Since the transition rules of BPA is source-dependent, and the transition rules for the opponent's alternative operator contain a fresh operator in their
sources, so GameBPA is a conservative extension of BPA.
\end{proof}

\begin{theorem}
Bisimulation equivalence is a congruence with respect to GameBPA.
\end{theorem}

\begin{proof}
The transition rules for the opponent's alternative operator, as well as of BPA, are all in panth format. So the bisimulation equivalence that they induce is a congruence.
\end{proof}

\subsection{Axioms of Opponent's Alternative Composition Operator}

By extending the opponent's alternative composition operator $\ddagger$ to model ACP for games, it is clear to construct a sound and complete axiomatic system. We design axioms of opponent's alternative composition operator $\ddagger$ shown in Table \ref{AxiomOfOAC}.

\begin{center}
\begin{table}
  \begin{tabular}{@{}ll@{}}
\hline No. &Axiom\\
  OA1 & $x, y\subset A_{\lambda_O}$, $x \ddagger y= x + y$ \\
\end{tabular}
\caption{Axioms of the opponent's alternative composition operator and the non-determinacy constant}
\label{AxiomOfOAC}
\end{table}
\end{center}

The axioms OA1 is presented for the opponent's alternative composition $\ddagger$.

\subsection{Properties of the Axiomatic System}

The following are two properties of the axiomatic system $\mathcal{E}_{\textrm{GameBPA}}$.

\begin{theorem}
$\mathcal{E}_{\textrm{GameBPA}}$  is sound for GameBPA modulo bisimulation equivalence.
\end{theorem}

\begin{proof}
Since bisimulation is both an equivalence and a congruence, we only need to check that the first clause in the definition of the relation $=$ is sound.
That is, if $s=t$ is an axiom in GameBPA and $\sigma$ is a closed substitution that maps the variables in $s$ and $t$ to process terms, then we need to check that $\sigma(s)\underline{\leftrightarrow}\sigma(t)$.

We only provide some intuition for soundness of the axioms in Table \ref{AxiomOfOAC}.

\begin{enumerate}
  \item The axiom OA1 says a GameBPA process term $t \ddagger s$ is the same as the process term $t + s$ in the view of O.
\end{enumerate}
\end{proof}

These intuitions can be made rigorous by means of explicit bisimulation relations between the left- and right-hand sides of closed instantiations of the axioms in Table \ref{AxiomOfOAC}. Hence, all such instantiations are sound modulo bisimulation equivalence.

\begin{theorem}
$\mathcal{E}_{\textrm{GameBPA}}$ is complete for GameBPA modulo bisimulation equivalence.
\end{theorem}

\begin{proof}
The proof is based on the proof of the completeness theorem of BPA. (See Section \ref{ACPP} and \cite{ACP}).

The proof consists of three main step: (1) we will show that the axioms OA1 can be turned in to rewrite rules, and the resulting TRS is terminating; (2) we will show that norm forms do not contain occurrences of the fresh opponent's alternative composition operator $\ddagger$; (3) we will prove that $\mathcal{E}_{\textrm{GameBPA}}$ is complete for GameBPA modulo bisimulation equivalence.

(1) The axioms OA1 is turned into rewriting rules directly from left to right, and added to the rewriting rule in the proof the completeness of $\mathcal{E}_{\textrm{BPA}}$ (see \cite{ACP}). The resulting TRS is terminating modulo AC (Associativity and Commutativity) of $+$ operator through defining new weight functions on process terms.

$$
weight(s\ddagger t)\triangleq weight(s) + weight(t)
$$

We can get that each application of a rewriting rule strictly decreases the weight of a process term, and that moreover process terms that are equivalent modulo AC of + have the same weight. Hence, the TRS is terminating modulo AC of $+$.

(2)We will show that the normal form $n$ are not of the form $s\ddagger t$. The proof is based on induction with respect to the size of the normal form $n$.

\begin{itemize}
  \item If n is an atomic action, then it does not contain $\ddagger$.
  \item Suppose $n =_{AC} s + t$ or $n =_{AC} s \cdot t$. Then by induction, the normal forms $s$ and $t$ do not contain $\ddagger$, so $n$ does not contain $\ddagger$.
  \item $n$ cannot be of the form $s \ddagger t$, because in that case, the directed version of OA1 would apply to it, contradicting the fact that $n$ is a normal form.
\end{itemize}

We proved that normal forms are all basic process terms.

(3)We proceed to prove that the axiomatization $\mathcal{E}_{\textrm{GameBPA}}$ is complete for GameBPA modulo bisimulation equivalence. Let the process terms $s$ and $t$ be bisimilar. The TRS is terminating modulo AC of the $+$, so it reduces $s$ and $t$ to normal forms $n$ and $n'$, respectively. Since the rewrite rules and equivalence modulo AC of the + can be derived from $\mathcal{E}_{\textrm{GameBPA}}$, $s = n$ and $t = n'$. Soundness of $\mathcal{E}_{\textrm{GameBPA}}$ then yields $s\underline{\leftrightarrow} n$ and $t \underline{\leftrightarrow} n'$, so $n\underline{\leftrightarrow} s \underline{\leftrightarrow} t \underline{\leftrightarrow} n'$. We shown that the normal forms $n$ and $n'$ are basic process terms. Then it follows that $n\underline{\leftrightarrow} n'$ implies $n =_{AC} n'$. Hence, $s = n =_{AC} n' = t$.
\end{proof}

\subsection{Execution of GameBPA Processes}

Execution of GameBPA processes needs the co-action of the P and the O, that is, the playing of the game between the P and the O. we introduce a new binary playing operator $\sqcap$ over the P's GameBPA process and the O's GameBPA process. A single GameBPA process is non-deterministic, but the playing operator $\sqcap$ of GameBPA processes for the P and the O will eliminate this determinacy and will result in a real execution of two GameBPA processes. To eliminate the mismatched branches in the alternation composition in the co-action of GameBPA processes, a special constant called deadlock $\delta$ is also introduced.

\subsubsection{Transition Rules of Playing Operator}

The transition rules of playing operator $\sqcap$ are following.

$$
\frac{x\xrightarrow{\upsilon}\surd \quad y\xrightarrow{\upsilon}\surd}{x\sqcap y\xrightarrow{\upsilon}\surd}
$$

$$
\frac{x\xrightarrow{\upsilon}\surd \quad y\xrightarrow{\upsilon}y'}{x\sqcap y\xrightarrow{\upsilon}y'}
$$

$$
\frac{x\xrightarrow{\upsilon}x' \quad y\xrightarrow{\upsilon}\surd}{x\sqcap y\xrightarrow{\upsilon}x'}
$$

$$
\frac{x\xrightarrow{\upsilon}x' \quad y\xrightarrow{\upsilon}y'}{x\sqcap y\xrightarrow{\upsilon}x'\sqcap y'}
$$

The first transition rule of playing operator $\sqcap$ says if $t$ terminates successfully by executing an action $a$ and $s$ terminates successfully by executing the same action $a$, then the process term $t \sqcap s$ will terminate successfully by executing the action $a$. The second one says if $t$ terminates successfully by executing an action $a$ and $s$ evolves into $s'$ by executing the same action $a$, then the process term $t \sqcap s$ will evolve into $s'$ by executing the same action $a$. The third one says if $t$ evolves into $t'$ by executing an action $a$ and $s$ terminates successfully by executing the same action $a$, then the process term $t \sqcap s$ will evolve into $t'$ by executing the same action $a$. The fourth one says if $t$ evolves into $t'$ by executing an action $a$ and $s$ evolves into $s'$ by executing the same action $a$, then the process term $t \sqcap s$ will evolve into $t' \sqcap s'$ by executing the same action $a$.

The above four transition rules intuitively capture the co-action of the P's GameBPA process and the O's GameBPA process.

To eliminate the mismatched branches in the alternation composition in the co-action of GameBPA processes, we introduce a special constant called deadlock $\delta$. The deadlock $\delta$ means do nothing. That is, when the execution sequence of the P's process is not matched that of the O's process, a deadlock will be caused.

\subsubsection{Properties of Playing Operator}

\begin{theorem}
GameBPA with playing operator and deadlock constant is a conservative extension of GameBPA.
\end{theorem}

\begin{proof}
This theorem follows from the following two facts.

\begin{enumerate}
  \item The transition rules of GameBPA are all source-dependent.
  \item The sources of the four transition rules for the playing operator all contain an occurrence of $\sqcap$.
\end{enumerate}

Since the transition rules of GameBPA is source-dependent, and the transition rules for the playing operator contain a fresh operator in their
sources, so GameBPA with playing operator is a conservative extension of GameBPA.
\end{proof}

\begin{theorem}
Bisimulation equivalence is a congruence with respect to GameBPA with playing operator and deadlock constant.
\end{theorem}

\begin{proof}
The transition rules for the playing operator, as well as of GameBPA, are all in panth format. So the bisimulation equivalence that they induce is a congruence.
\end{proof}

\subsubsection{Axioms of Playing Operator and Deadlock Constant}

We design the axioms of the playing operator and the deadlock constant as Table \ref{AxiomOfPO} shows.

\begin{center}
\begin{table}
  \begin{tabular}{@{}ll@{}}
\hline No. &Axiom\\
  DL1 & $x + \delta = x$\\
  DL2 & $\delta\cdot x = \delta$\\
  PO1 & $\upsilon \sqcap \upsilon = \upsilon$ \\
  PO2 & $\upsilon \sqcap \omega = \delta$ \\
  PO3 & $\delta \sqcap x = \delta$ \\
  PO4 & $x \sqcap \delta = \delta$ \\
  PO5 & $\upsilon \sqcap (\upsilon\cdot y) = \upsilon\cdot y$ \\
  PO6 & $\upsilon \sqcap (\omega\cdot y) = \delta$\\
  PO7 & $(\upsilon \cdot x)\sqcap \upsilon = \upsilon\cdot x$\\
  PO8 & $(\upsilon \cdot x)\sqcap \omega = \delta$\\
  PO9 & $(\upsilon \cdot x)\sqcap (\upsilon\cdot y) = \upsilon\cdot(x\sqcap y)$\\
  PO10 & $(\upsilon \cdot x)\sqcap (\omega\cdot y) = \delta$\\
  PO11 & $(x\ddagger y)\sqcap z = (x + y)\sqcap z$\\
  PO12 & $x\sqcap (y\ddagger z) = x\sqcap (y + z)$\\
  PO13 & $(x+ y)\sqcap z = x\sqcap z + y\sqcap z$\\
  PO14 & $x\sqcap (y + z) = x\sqcap y + x\sqcap z$\\
\end{tabular}
\caption{Axioms of playing operator of GameBPA processes and deadlock constant}
\label{AxiomOfPO}
\end{table}
\end{center}

The axioms DL1-DL2 are presented for the deadlock constant $\delta$, and the axioms PO1-PO14 are for the playing operator $\sqcap$. There are not axioms for the association of the deadlock constant $\delta$ and the playing operator $\sqcap$, just because the function of the playing operator $\sqcap$ is eliminating all non-deterministic factors.

\subsubsection{Properties of the Axiomatic System}

\begin{theorem}
$\mathcal{E}_{\textrm{GameBPA}}$ + DL1-DL2 + PO1-PO14  is sound for GameBPA with playing operator and deadlock constant modulo bisimulation equivalence.
\end{theorem}

\begin{proof}
Since bisimulation is both an equivalence and a congruence, we only need to check that the first clause in the definition of the relation $=$ is sound.
That is, if $s=t$ is an axiom in GameBPA and $\sigma$ is a closed substitution that maps the variables in $s$ and $t$ to process terms, then we need to check that $\sigma(s)\underline{\leftrightarrow}\sigma(t)$.

We only provide some intuition for soundness of the axioms in Table \ref{AxiomOfPO}.

\begin{enumerate}
  \item The axiom DL1 says that $\delta$ displays no behavior, so the process term $t + \delta$ is equal to the process term $t$.
  \item The axioms DL2, PO3 and PO4 say that $\delta$ blocks the behavior of the process term $\delta \cdot t$, $\delta \sqcap t$ and $t \sqcap \delta$.
  \item The axioms PO1 and PO2 say that the co-action of two same actions will lead to the only action, otherwise, it will cause a deadlock.
  \item The axioms PO5-PO10 say that $s\sqcap t$ makes as initial transition a playing of initial transitions from $s$ and $t$. If the execution sequence of $s$ is not matched with that of $t$, a deadlock will be caused.
  \item The axioms PO11-PO12 say that the function of playing operator makes two non-deterministic GameBPA processes deterministic.
  \item The axioms PO13-PO14 say that the playing operator satisfies right and left distributivity to the operator $+$.
\end{enumerate}

These intuitions can be made rigorous by means of explicit bisimulation relations between the left- and right-hand sides of closed instantiations of the axioms in Table \ref{AxiomOfPO}. Hence, all such instantiations are sound modulo bisimulation equivalence.
\end{proof}

\begin{theorem}
$\mathcal{E}_{\textrm{GameBPA}}$ + DL1-DL2 + PO1-PO14 is complete for GameBPA with playing operator and deadlock constant modulo bisimulation equivalence.
\end{theorem}

\begin{proof}
The proof is based on the proof of the Theorem 4.

The proof consists of three main step: (1) we will show that the axioms DL1, DL2 and PO1-PO14 can be turned in to rewrite rules, and the resulting TRS is terminating; (2) we will show that norm forms do not contain occurrences of the fresh opponent's alternative composition operator $\sqcap$; (3) we will prove that $\mathcal{E}_{\textrm{GameBPA}}$ + DL1-DL2 + PO1-PO14 is complete for GameBPA with playing operator and deadlock constant modulo bisimulation equivalence.

(1) The axioms DL1-DL2 + PO1-PO14 is turned into rewriting rules directly from left to right, and added to the rewriting rules in the proof the completeness of $\mathcal{E}_{\textrm{GameBPA}}$ (see proof of Theorem 4). The resulting TRS is terminating modulo AC (Associativity and Commutativity) of $+$ operator through defining new weight functions on process terms.

$$
weight(\delta)\triangleq 2
$$

$$
weight(\upsilon)\triangleq 2
$$

$$
weight(\omega)\triangleq 2
$$

$$
weight(s\sqcap t)\triangleq (weight(s) \cdot weight(t))^2
$$

We can get that each application of a rewriting rule strictly decreases the weight of a process term, and that moreover process terms that are equivalent modulo AC of + have the same weight. Hence, the TRS is terminating modulo AC of $+$.

(2)We will show that the normal form $n$ are not of the form $s\sqcap t$. The proof is based on induction with respect to the size of the normal form $n$.

\begin{itemize}
  \item If n is an atomic action, then it does not contain $\sqcap$.
  \item Suppose $n =_{AC} s + t$ or $n =_{AC} s \cdot t$. Then by induction, the normal forms $s$ and $t$ do not contain $\sqcap$, so $n$ does not contain $\sqcap$.
  \item Suppose $n =_{AC} s\sqcap t$. By induction, the normal form $s$ does not contain $\sqcap$. We distinguish the possible forms of the normal form $s$:
  \begin{itemize}
    \item if $s \equiv a$, then the directed version of PO1, PO2, PO5 or PO6 apply to $s \sqcap t$;
    \item if $s =_{AC} au$, then the directed version of PO7-PO10 apply to $s\sqcap t$;
    \item if $s =_{AC} u + u'$, then the directed version of PO13 applies to $s \sqcap t$;
    \item if $s =_{AC} u \ddagger u'$, then the directed version of PO11 applies to $s \sqcap t$. (Actually, we already prove that $\ddagger$ cannot occur in the norm forms, see the proof of Theorem 4).
  \end{itemize}
   These four cases, which cover the possible forms of the normal form $s$, contradict the fact that $n$ is a normal form. Similarly, we can induce the possible forms of the normal form $t$. So, we conclude that $n$ cannot be of the form $s\sqcap t$.
\end{itemize}

We proved that normal forms are all basic process terms.

(3)We proceed to prove that the axiomatization $\mathcal{E}_{\textrm{GameBPA}}$ + DL1-DL2 + PO1-PO14 is complete for GameBPA with playing operator and deadlock constant modulo bisimulation equivalence. Let the process terms $s$ and $t$ be bisimilar. The TRS is terminating modulo AC of the $+$, so it reduces $s$ and $t$ to normal forms $n$ and $n'$, respectively. Since the rewrite rules and equivalence modulo AC of the + can be derived from $\mathcal{E}_{\textrm{GameBPA}}$ + DL1-DL2 + PO1-PO14, $s = n$ and $t = n'$. Soundness of $\mathcal{E}_{\textrm{GameBPA}}$ + DL1-DL2 + PO1-PO14 then yields $s\underline{\leftrightarrow} n$ and $t \underline{\leftrightarrow} n'$, so $n\underline{\leftrightarrow} s \underline{\leftrightarrow} t \underline{\leftrightarrow} n'$. We shown that the normal forms $n$ and $n'$ are basic process terms. Then it follows that $n\underline{\leftrightarrow} n'$ implies $n =_{AC} n'$. Hence, $s = n =_{AC} n' = t$.
\end{proof}

\section{GameACP -- A Full Extension of ACP for Games}

GameBPA extends to process algebra BPA and does not use the full outcomes of ACP, such as concurrency, recursion, abstraction, etc. Now, we make GameBPA be based on the full ACP (exactly $\textrm{ACP}_\tau$ with guarded linear recursion) and this extension is called GameACP. GameACP remains the opponent's alternative composition operator $\ddagger$, the playing operator $\sqcap$. Because the deadlock constant is already existing in ACP, we remove the duplicate definition of deadlock constant in GameACP.

The transition rules of the opponent's alternative composition operator $\ddagger$ and the playing operator $\sqcap$ are the same as those in GameBPA. Through defining $\upsilon \cdot (\tau\cdot(x \ddagger y) \ddagger x) = \upsilon \cdot (x \ddagger y)$ , we extend $A$ to $A\cup \{\tau\}$.

We can get the following two conclusions.

\begin{theorem}
GameACP (exactly $\textrm{ACP}_\tau$ with guarded linear recursion, opponent's alternative composition operator $\ddagger$, playing operator $\sqcap$ is a conservative extension of ACP (exactly $\textrm{ACP}_\tau$ with guarded linear recursion) (see Section \ref{ACPP}).
\end{theorem}

\begin{proof}
The sources of transition rules of opponent's alternative composition operator $\ddagger$ and playing operator $\sqcap$ contain one fresh function symbol $\ddagger$ and $\sqcap$. And it is known that the transition rules of $\textrm{ACP}_\tau$ with guarded linear recursion are source-dependent. According to the definition of conservative extension, GameACP is a conservative extension of $\textrm{ACP}_\tau$ with guarded linear recursion.
\end{proof}

\begin{theorem}
Rooted branching bisimulation equivalence is a congruence with respect to GameACP (exactly $\textrm{ACP}_\tau$ with guarded linear recursion, opponent's alternative composition operator $\ddagger$, playing operator $\sqcap$.
\end{theorem}

\begin{proof}
We introduce successful termination predicate $\downarrow$. A transition rule $\frac{}{\surd\downarrow}$ is added into transition rules of GameACP. Replacing transition rules occurring $\xrightarrow{a}\surd$ by $\xrightarrow{a}\surd\downarrow$, the result transition rules of GameACP are in RBB cool format according to the definition of RBB cool format. So rooted branching bisimulation equivalence is a congruence with respect to GameACP according to the definition of congruence.
\end{proof}

Because of the remove of the deadlock constant in GameACP, the axiomatization $\mathcal{E}_{\textrm{GameACP}}$ of GameACP (exactly $\textrm{ACP}_\tau$ with guarded linear recursion, opponent's alternative composition operator $\ddagger$, playing operator $\sqcap$) only contains $\mathcal{E}_{\textrm{ACP}_\tau}$ + RDP, RSP, CFAR and OA1, PO1-PO14.

Now, we get the following two conclusions.

\begin{theorem}
$\mathcal{E}_{\textrm{GameACP}}$ ($\mathcal{E}_{\textrm{ACP}_\tau}$ + RDP, RSP, CFAR + OA1 + PO1-PO14) is sound for GameACP (exactly $\textrm{ACP}_\tau$ with guarded linear recursion, opponent's alternative composition operator $\ddagger$, playing operator $\sqcap$) modulo rooted branching bisimulation equivalence.
\end{theorem}

\begin{proof}
Because rooted branching bisimulation is both an equivalence and a congruence, we only need to check that if $t=u$ is an axiom and a closed substitution $\sigma$ replacing the variables in $t$ and $u$ to get $\sigma(t)$ and $\sigma(u)$, then $\sigma(t)\underline{\leftrightarrow}_{rb}\sigma(u)$.

We only provide some intuition for soundness of the axioms AO1, PO1-PO14.

\begin{enumerate}
  \item The axiom OA1 says a GameBPA process term $t \ddagger s$ is the same as the process term $t + s$ in the view of O.
  \item The axioms PO3 and PO4 say that $\delta$ blocks the behavior of the process term $\delta \cdot t$, $\delta \sqcap t$ and $t \sqcap \delta$.
  \item The axioms PO1 and PO2 say that the co-action of two same actions will lead to the only action, otherwise, it will cause a deadlock.
  \item The axioms PO5-PO10 say that $s\sqcap t$ makes as initial transition a playing of initial transitions from $s$ and $t$. If the execution sequence of $s$ is not matched with that of $t$, a deadlock will be caused.
  \item The axioms PO11-PO12 say that the function of playing operator makes two non-deterministic GameBPA processes deterministic.
  \item The axioms PO13-PO14 say that the playing operator satisfies right and left distributivity to the operator $+$.
\end{enumerate}

These intuitions can be made rigorous by means of explicit rooted branching bisimulation relations between the left- and right-hand sides of closed instantiations of the axioms in Table \ref{AxiomOfPO}. Hence, all such instantiations are sound modulo rooted branching bisimulation equivalence.
\end{proof}

\begin{theorem}
$\mathcal{E}_{\textrm{GameACP}}$ ($\mathcal{E}_{\textrm{ACP}_\tau}$ + RDP, RSP, CFAR + OA1 + PO1-PO14) is complete for GameACP (exactly $\textrm{ACP}_\tau$ with guarded linear recursion, opponent's alternative composition operator $\ddagger$, playing operator $\sqcap$) modulo rooted branching bisimulation equivalence.
\end{theorem}

\begin{proof}
We need to prove that each process term $t$ in GameACP is equal to a process term $\langle X|E\rangle$ with a guarded linear recursive specification $E$. That is, if $\langle X_1|E_1\rangle\underline{\leftrightarrow}_{rb}\langle Y_1|E_2\rangle$ for guarded linear recursive specifications $E_1$ and $E_2$, then $\langle X_1|E_1\rangle=\langle Y_1|E_2\rangle$ can be gotten from $\mathcal{E}_{\textrm{GameACP}}$.

This proof is based on the completeness proof\cite{ACP} of $\mathcal{E}_{\textrm{ACP}_\tau}$ + RDP, RSP, CFAR. We apply structural induction the size of process term $t$. The new case is $t\equiv s\sqcap r$. First assuming $s=\langle X_1|E\rangle$ with a guarded linear recursive specification $E$ and $r=\langle Y_1|F\rangle$ with a guarded linear recursive specification $F$, we prove the case of $t=\langle X_1|E\rangle\sqcap \langle Y_1|F\rangle$. Let $E$ consists of guarded linear recursive equations

$$X_i=a_{i1}X_{i1}+...+a_{ik_i}X_{ik_i}+b_{i1}+...+b_{il_i}$$
for $i\in{1,...,N}$. Let $F$ consists of guarded linear recursive equations

$$Y_j=c_{j1}Y_{j1}+...+c_{jm_j}Y_{jm_j}+d_{j1}+...+d_{jn_j}$$
for $j\in{1,...,M}$.

\begin{eqnarray}
&&\langle X_i|E\rangle\sqcap \langle Y_j|F\rangle \nonumber\\
&\overset{\text{RDP}}{=}&(a_{i1}X_{i1}+...+a_{ik_i}X_{ik_i}+b_{i1}+...+b_{il_i})\sqcap \langle Y_j|F\rangle \nonumber\\
&\overset{\text{PO13}}{=}&a_{i1}X_{i1}\sqcap \langle Y_j|F\rangle +...+a_{ik_i}X_{ik_i}\sqcap  \langle Y_j|F\rangle \nonumber\\
&&+b_{i1}\sqcap \langle Y_j|F\rangle +...+b_{il_i}\sqcap \langle Y_j|F\rangle \nonumber\\
&\overset{\text{RDP}}{=}&a_{i1}X_{i1}\sqcap (c_{j1}Y_{j1}+...+c_{jm_j}Y_{jm_j}+d_{j1}+...+d_{jn_j})\nonumber\\
&&+...+a_{ik_i}X_{ik_i}\sqcap  (c_{j1}Y_{j1}+...+c_{jm_j}Y_{jm_j}+d_{j1}+...+d_{jn_j}) \nonumber\\
&&+b_{i1}\sqcap (c_{j1}Y_{j1}+...+c_{jm_j}Y_{jm_j}+d_{j1}+...+d_{jn_j})\nonumber\\
&&+...+b_{il_i}\sqcap (c_{j1}Y_{j1}+...+c_{jm_j}Y_{jm_j}+d_{j1}+...+d_{jn_j}) \nonumber\\
&\overset{\text{PO14}}{=}&(a_{i1}X_{i1})\sqcap (c_{j1}Y_{j1})+...+(a_{i1}X_{i1})\sqcap (c_{jm_j}Y_{jm_j})+(a_{i1}X_{i1})\sqcap (d_{j1})+...+(a_{i1}X_{i1})\sqcap (d_{jn_j})\nonumber\\
&&+...+(a_{ik_i}X_{ik_i})\sqcap (c_{j1}Y_{j1})+...+(a_{ik_i}X_{ik_i})\sqcap (c_{jm_j}Y_{jm_j})+(a_{ik_i}X_{ik_i})\sqcap (d_{j1})+...+(a_{ik_i}X_{ik_i})\sqcap (d_{jn_j})\nonumber\\
&&+(b_{i1})\sqcap (c_{j1}Y_{j1})+...+(b_{i1})\sqcap (c_{jm_j}Y_{jm_j})+(b_{i1})\sqcap (d_{j1})+...+(b_{i1})\sqcap (d_{jn_j})\nonumber\\
&&+...+(b_{il_i})\sqcap (c_{j1}Y_{j1})+...+(b_{il_i})\sqcap (c_{jm_j}Y_{jm_j})+(b_{il_i})\sqcap (d_{j1})+...+(b_{il_i})\sqcap (d_{jn_j})\nonumber
\end{eqnarray}

Then we can use the axioms PO1-PO10 into the above equation. This will lead to several cases and we do not enumerate all these cases. But, we can see that every case will lead to a guarded linear recursive specification.
\end{proof}

\section{GameAPTC -- A Full Extension of APTC for Games}

GameACP is an extension for games based on ACP, but there is a problem. Assuming that $P=a+b$ and $O=c+d$, then the whole system in parallel is $P\parallel O=(a+b)\parallel (c\ddagger d)$. We know that $P\parallel O=P\cdot O + O\cdot P$, a new non-deterministic operator $+$ is occurred, and we can not determine if it is belonging to $P$ or $O$, that is, $P\parallel O=(a+b)\cdot(c\ddagger d)+(c\ddagger d)\cdot(a+b)$ or $P\parallel O=(a+b)\cdot(c\ddagger d)\ddagger(c\ddagger d)\cdot(a+b)$ can not be determined.

To avoid the new non-determinacy, it is needed to extended for games based on truly concurrent process algebra APTC \cite{APTC}, because the parallelism in APTC can not introduce new kind of non-determinacy. Since the axiom system of BATC for games is the same as that of BPA for games, it is only need to prove that the axioms of new operators $\ddagger$ and $\sqcap$ are sound and complete modulo truly concurrent bisimilarities $\sim_p$, $\sim_s$, $\sim_{hp}$ and $\sim_{hhp}$, we omit them and directly extend APTC for games.

The axioms of $\ddagger$ and $\sqcap$ are shown in Table \ref{AxiomOfPOAPTC}.

\begin{center}
\begin{table}
  \begin{tabular}{@{}ll@{}}
  \hline No. &Axiom\\
  PO1 & $\upsilon \sqcap \upsilon = \upsilon$ \\
  PO2 & $\upsilon \sqcap \omega = \delta$ \\
  PO3 & $\delta \sqcap x = \delta$ \\
  PO4 & $x \sqcap \delta = \delta$ \\
  PO5 & $\upsilon \sqcap (\upsilon\cdot y) = \upsilon\cdot y$ \\
  PO6 & $\upsilon \sqcap (\omega\cdot y) = \delta$\\
  PO7 & $(\upsilon \cdot x)\sqcap \upsilon = \upsilon\cdot x$\\
  PO8 & $(\upsilon \cdot x)\sqcap \omega = \delta$\\
  PO9 & $(\upsilon \cdot x)\sqcap (\upsilon\cdot y) = \upsilon\cdot(x\sqcap y)$\\
  PO10 & $(\upsilon \cdot x)\sqcap (\omega\cdot y) = \delta$\\
  PO11 & $(x\ddagger y)\sqcap z = (x + y)\sqcap z$\\
  PO12 & $x\sqcap (y\ddagger z) = x\sqcap (y + z)$\\
  PO13 & $(x+ y)\sqcap z = x\sqcap z + y\sqcap z$\\
  PO14 & $x\sqcap (y + z) = x\sqcap y + x\sqcap z$\\
  PO15 & $(x\ddagger y)\parallel z = x\parallel z\ddagger y\parallel z$\\
  PO16 & $x\parallel (y\ddagger z)=x\parallel y\ddagger x\parallel z$\\
  PO17 & $(x\parallel y)\sqcap z = (x\sqcap z)\parallel (y\sqcap z)$\\
  PO18 & $x\sqcap(y\parallel z)=(x\sqcap y)\parallel (y\sqcap z)$\\
\end{tabular}
\caption{Axioms of playing operator based on APTC}
\label{AxiomOfPOAPTC}
\end{table}
\end{center}

We can get the following conclusions, and the proofs are omitted.

\begin{theorem}
GameAPTC (exactly $\textrm{APTC}_\tau$ with guarded linear recursion, opponent's alternative composition operator $\ddagger$, playing operator $\sqcap$ is a conservative extension of APTC (exactly $\textrm{APTC}_\tau$ with guarded linear recursion).
\end{theorem}

\begin{theorem}
Rooted branching truly concurrent bisimulation equivalences $\sim_p$, $\sim_s$ and $\sim_{hp}$ are all congruences with respect to GameAPTC.
\end{theorem}

\begin{theorem}
$\mathcal{E}_{\textrm{GameAPTC}}$ ($\mathcal{E}_{\textrm{APTC}_\tau}$ + RDP, RSP, CFAR + OA1 + PO1-PO18) is sound for GameACP modulo rooted branching truly concurrent bisimulation equivalences $\sim_p$, $\sim_s$ and $\sim_{hp}$.
\end{theorem}

\begin{theorem}
$\mathcal{E}_{\textrm{GameAPTC}}$ ($\mathcal{E}_{\textrm{APTC}_\tau}$ + RDP, RSP, CFAR + OA1 + PO1-PO18) is complete for GameACP modulo rooted branching truly concurrent bisimulation equivalences $\sim_p$, $\sim_s$ and $\sim_{hp}$.
\end{theorem}

\section{Correctness Theorem}

\begin{theorem}
If $\lambda_P$ is a P strategy and $\lambda_O$ is an O strategy as illustrated in Section \ref{Games}, and if the GameACP or GameAPTC process term $s$ corresponds to $\lambda_P$ and the GameACP process term $t$ corresponds to $\lambda_O$, then the process term $s \sqcap t$ exactly defines an execution of $\lambda_P$ and $\lambda_O$.
\end{theorem}

\begin{proof}
We will show that $s\sqcap t$ exactly results in the maximal element of $\lambda_P \cap \lambda_O$.

The axioms PO11, PO12 make non-deterministic GameACP or GameAPTC processes deterministic.

The axioms PO13, PO14 inspect all deterministic branches.

The axioms DL2, PO2, PO3, PO4, PO6, PO8, PO10 assure that the mismatched execution sequence will cause a deadlock.

The axiom DL1 eliminates the deadlock branches in a GameACP or GameAPTC process.

The axioms PO1, PO5, PO7 and PO9 assure occurrence of the matched execution sequence of two GameACP or GameAPTC processes.

The axioms PO5, PO7 assure the selection the maximal execution sequence.
\end{proof}

We illustrate the correctness theorem through three examples in Section \ref{AS}.

For the P strategy$\lambda_P$ corresponding to the GameACP or GameAPTC process term $start\cdot write\cdot (submit\cdot store \ddagger cancel)$ and the O strategy corresponding to the GameACP or GameAPTC process term $start\cdot write\cdot submit\cdot store$ in Fig. \ref{SubmittingOrder}, the maximal element $start\cdot write\cdot submit\cdot store$ of $\lambda_P \cap \lambda_O$ defines an execution of  process graph Fig. \ref{SubmittingOrder}-b.

\begin{eqnarray}
&&(start\cdot write\cdot (submit\cdot store \ddagger cancel))\sqcap (start\cdot write\cdot submit\cdot store)\nonumber\\
&\overset{\text{PO9}}{=}&start\cdot write \cdot((submit\cdot store \ddagger cancel)\sqcap(submit\cdot store))\nonumber\\
&\overset{\text{PO11}}{=}&start\cdot write \cdot((submit\cdot store + cancel)\sqcap (submit\cdot store))\nonumber\\
&\overset{\text{PO13,PO6}}{=}&start\cdot write \cdot (submit\cdot store + \delta)\nonumber\\
&\overset{\text{DL1}}{=}&start\cdot write \cdot submit\cdot store\nonumber
\end{eqnarray}

For the P strategy $\lambda_P$ corresponding to the GameACP or GameAPTC process term $start\cdot operate\cdot (submit\cdot store \ddagger abort\cdot rollback)$ and the O strategy corresponding to the GameACP or GameAPTC process term $start\cdot operate\cdot abort\cdot rollback$ in Fig. \ref{Transaction}, the maximal element $start\cdot operate\cdot abort\cdot rollback$ of $\lambda_P \cap \lambda_O$ defines an execution of the process as illustrated in process graph Fig. \ref{Transaction}-b.

\begin{eqnarray}
&&(start\cdot operate\cdot (submit\cdot store \ddagger abort\cdot rollback))\sqcap (start\cdot operate\cdot abort\cdot rollback)\nonumber\\
&\overset{\text{PO9}}{=}&start\cdot operate\cdot((submit\cdot store \ddagger abort\cdot rollback)\sqcap (abort \cdot rollback))\nonumber\\
&\overset{\text{PO11}}{=}&start\cdot operate\cdot((submit\cdot store + abort \cdot rollback)\sqcap (abort \cdot rollback))\nonumber\\
&\overset{\text{PO13,PO6}}{=}&start\cdot operate\cdot(\delta + abort\cdot rollback)\nonumber\\
&\overset{\text{DL1}}{=}&start\cdot operate\cdot abort\cdot rollback\nonumber
\end{eqnarray}

For the P strategy $\lambda_P$ corresponding to the GameACP or GameAPTC process term $start\cdot shopping\cdot (sTruck\cdot oTruck\cdot pOnLine \ddagger sTrain\cdot oTrain\cdot pOnLine \ddagger sPlane \cdot oPlane \cdot pOffLine)$ in Fig. \ref{PurchasingPlayer} and the O strategy corresponding to the GameACP or GameAPTC process term $start\cdot shopping\cdot sPlane \cdot oPlane \cdot (pOffLine \ddagger pOnLine)$ in Fig. \ref{PurchasingOpponent}, the maximal element $start\cdot shopping\cdot sPlane \cdot oPlane \cdot pOffLine$ of $\lambda_P \cap \lambda_O$ defines an execution of the process as illustrated in Fig. \ref{Execution}.

\begin{eqnarray}
&&(start\cdot shopping\cdot (sTruck\cdot oTruck\cdot pOnLine \ddagger sTrain\cdot oTrain\cdot pOnLine\nonumber\\
&&\ddagger sPlane \cdot oPlane \cdot pOffLine))\nonumber\\
&&\sqcap (start\cdot shopping\cdot sPlane \cdot oPlane \cdot (pOffLine \ddagger pOnLine))\nonumber\\
&\overset{\text{PO9}}{=}&start\cdot shopping\cdot(sTruck\cdot oTruck\cdot pOnLine \ddagger sTrain\cdot oTrain\cdot pOnLine\nonumber\\
&&\ddagger sPlane \cdot oPlane \cdot pOffLine))\nonumber\\
&&\sqcap (sPlane \cdot oPlane \cdot (pOffLine \ddagger pOnLine)))\nonumber\\
&\overset{\text{PO11}}{=}&start\cdot shopping\cdot(sTruck\cdot oTruck\cdot pOnLine + sTrain\cdot oTrain\cdot pOnLine\nonumber\\
&&+ sPlane \cdot oPlane \cdot pOffLine))\nonumber\\
&&\sqcap (sPlane \cdot oPlane \cdot (pOffLine \ddagger pOnLine))\nonumber\\
&\overset{\text{PO13,PO10,PO9}}{=}&start\cdot shopping\cdot(\delta + \delta + sPlane \cdot oPlane \cdot (pOffLine \sqcap(pOffLine \ddagger pOnLine)))\nonumber\\
&\overset{\text{DL1,PO5,PO12}}{=}&start\cdot shopping\cdot(sPlane \cdot oPlane \cdot (pOffLine \sqcap(pOffLine + pOnLine)))\nonumber\\
&\overset{\text{PO14,PO2,PO1}}{=}&start\cdot shopping\cdot(sPlane \cdot oPlane \cdot (pOffLine+\delta))\nonumber\\
&\overset{\text{DL1}}{=}&start \cdot shopping \cdot sPlane \cdot oPlane \cdot pOffLine \nonumber
\end{eqnarray}

\section{Support for Multi-person Games -- Extended Purchasing Example}

In fact, the axioms in Table \ref{AxiomOfOAC} and Table \ref{AxiomOfPO} can be naturally used in multi-person games without any alternation. For a three-person games, let $t$ be a GameACP or GameAPTC process term corresponding to a strategy of the first player, $s$ a GameACP or GameAPTC process term corresponding to a strategy of the second player and $u$ a GameACP or GameAPTC process term corresponding to a strategy of the third player. The process term $t\sqcap s\sqcap u$ can be deduced to an execution of these strategies by use of the above axioms. We show this situation in the section.

\begin{figure}
  \centering
  %\vspace{5cm}
  \includegraphics{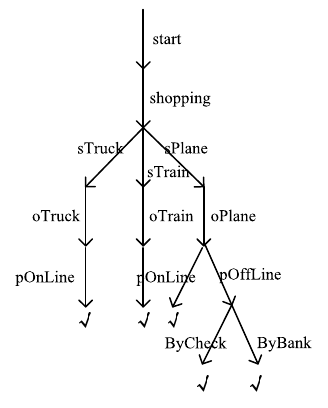}
  \caption{Process Graph of Purchasing Goods Example.}
  \label{Purchasing21}
\end{figure}

The process graph of the extended Purchasing composite Web Service is illustrated in Fig. \ref{Purchasing21}. The process of the Purchasing composite Web Service can be expressed by the following process term in ACP.

$start\cdot shopping\cdot (sTruck\cdot oTruck\cdot pOnLine + sTrain\cdot oTrain\cdot pOnLine + sPlane \cdot oPlane \cdot (pOnLine + pOffLine \cdot (ByCheck + ByBank)))$.

\begin{figure}
  \centering
  %\vspace{5cm}
  \includegraphics{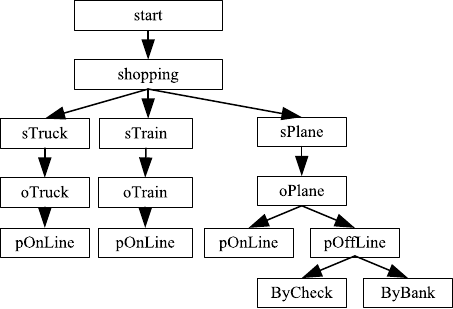}
  \caption{Game Tree of Purchasing Goods Example.}
  \label{PurchasingGameTree2}
\end{figure}

The game tree to process graph in Fig. \ref{Purchasing21} is illustrated in Fig. \ref{PurchasingGameTree2}.

In the game tree illustrated in Fig. \ref{PurchasingGameTree2} of the extended Purchasing example, there are three choice decisions. The first is made by the user agent (or the user), and the second is made by the composite service, and the third is made by the air corporation. In this game, we model the user agent as Player 1, the composite service as Player 2 and the air corporation as Player 3.

\begin{figure}
  \centering
  %\vspace{5cm}
  \includegraphics{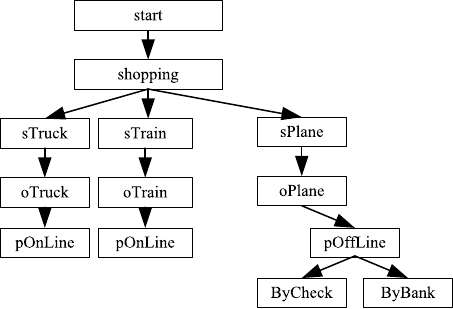}
  \caption{Player1 Strategy of Purchasing Game.}
  \label{PurchasingPlayer1}
\end{figure}

\begin{figure}
  \centering
  %\vspace{5cm}
  \includegraphics{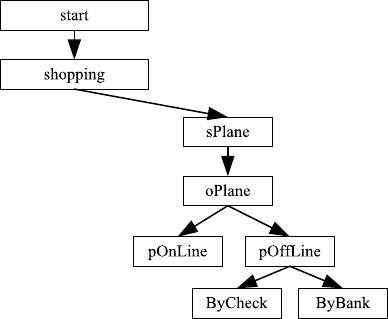}
  \caption{Player2 Strategy of Purchasing Game.}
  \label{PurchasingPlayer2}
\end{figure}

\begin{figure}
  \centering
  %\vspace{5cm}
  \includegraphics{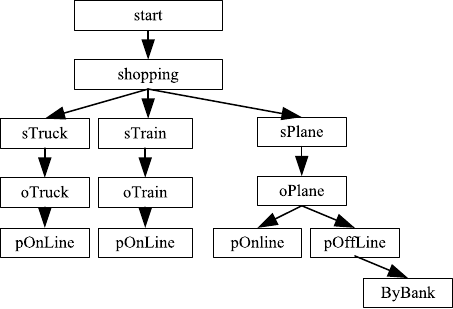}
  \caption{Player3 Strategy of Purchasing Game.}
  \label{PurchasingPlayer3}
\end{figure}

A strategy of Player 1 is illustrated as Fig. \ref{PurchasingPlayer1} shows. And a strategy of Player 2 is as Fig. \ref{PurchasingPlayer2} illustrates. And also Fig. \ref{PurchasingPlayer3} shows a strategy of Player 3.

We can see that the actual execution a game tree are acted together by all players. For a strategy $\lambda_{P1}$, a strategy $\lambda_{P2}$ and a strategy $\lambda_{P3}$ of a game tree, $\lambda_{P1} \cap \lambda_{P2} \cap \lambda_{P3}$ has the form $\lambda_{P1} \cap \lambda_{P2} \cap \lambda_{P3}=\{\epsilon,m_1,m_1\cdot m_2,...,m_1\cdot...\cdot m_n\}$ according to the definition of strategy. We can get that the maximal element $m_1\cdot...\cdot m_n$ of $\lambda_{P1} \cap \lambda_{P2} \cap \lambda_{P3}$ exactly defines an execution of the game tree.

\begin{figure}
  \centering
  %\vspace{5cm}
  \includegraphics{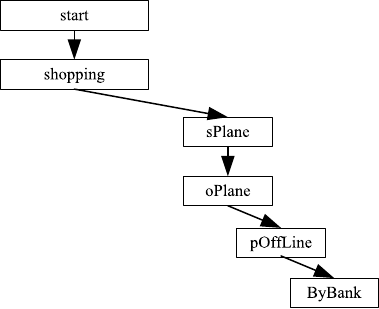}
  \caption{Execution of Purchasing Process.}
  \label{Execution2}
\end{figure}

For the strategy $\lambda_{P1}$ illustrated in Fig. \ref{PurchasingPlayer1}, the strategy $\lambda_{P2}$ illustrated in Fig. \ref{PurchasingPlayer2}, and the strategy $\lambda_{P3}$ illustrated in Fig. \ref{PurchasingPlayer3} the maximal element $start\cdot shopping\cdot sPlane \cdot oPlane \cdot pOffLine \cdot ByBank$ of $\lambda_{P1} \cap \lambda_{P2} \cap \lambda_{P3}$ defines an execution of the process as illustrated in process graph Fig. \ref{Purchasing2}. This is shown in Fig. \ref{Execution2}.

In extended Purchasing example, in the view of the Player 1, the process can be expressed by the following process term in GameACP or GameAPTC.

$start\cdot shopping\cdot (sTruck\cdot oTruck\cdot pOnLine + sTrain\cdot oTrain\cdot pOnLine + sPlane \cdot oPlane \cdot (pOnLine \ddagger pOffLine \cdot (ByCheck \ddagger ByBank)))$.

So the subtrees corresponding to term $start\cdot shopping\cdot sTruck\cdot oTruck\cdot pOnLine$, $start\cdot shopping\cdot sTrain\cdot oTrain\cdot pOnLine$ and $start\cdot shopping\cdot sPlane \cdot oPlane \cdot (pOnLine \ddagger pOffLine \cdot(ByCheck \ddagger ByBank))$ are all strategies of the Player 1.

In the view of the Player 2, the process can be expressed by the following process term in GameACP or GameAPTC.

$start\cdot shopping\cdot (sTruck\cdot oTruck\cdot pOnLine \ddagger sTrain\cdot oTrain\cdot pOnLine \ddagger sPlane \cdot oPlane \cdot (pOnLine + pOffLine \cdot (ByCheck \ddagger ByBank)))$.

So the subtrees corresponding to term $start\cdot shopping\cdot (sTruck\cdot oTruck\cdot pOnLine \ddagger sTrain\cdot oTrain\cdot pOnLine \ddagger sPlane \cdot oPlane \cdot pOnLine)$ and $start\cdot shopping\cdot (sTruck\cdot oTruck\cdot pOnLine \ddagger sTrain\cdot oTrain\cdot pOnLine \ddagger sPlane \cdot oPlane \cdot pOffLine \cdot (ByCheck \ddagger ByBank))$ are all strategies of the Player 2.

In the view of the Player 3, the process can be expressed by the following process term in GameACP or GameAPTC.

$start\cdot shopping\cdot (sTruck\cdot oTruck\cdot pOnLine \ddagger sTrain\cdot oTrain\cdot pOnLine \ddagger sPlane \cdot oPlane \cdot (pOnLine \ddagger pOffLine \cdot (ByCheck + ByBank)))$.

So the subtrees corresponding to term $start\cdot shopping\cdot (sTruck\cdot oTruck\cdot pOnLine \ddagger sTrain\cdot oTrain\cdot pOnLine \ddagger sPlane \cdot oPlane \cdot (pOnLine \ddagger pOffLine \cdot ByCheck))$ and $start\cdot shopping\cdot (sTruck\cdot oTruck\cdot pOnLine \ddagger sTrain\cdot oTrain\cdot pOnLine \ddagger sPlane \cdot oPlane \cdot (pOnLine \ddagger pOffLine \cdot ByBank))$ are all strategies of the Player 3.

If $\lambda_{P1}$ is a strategy of the Player 1, $\lambda_{P2}$ is a strategy of Player 2 and $\lambda_{P3}$ is a strategy of Player 3 as illustrated in Section \ref{Games}, and if the GameACP or GameAPTC process term $s$ corresponds to $\lambda_{P1}$, the GameACP or GameAPTC process term $t$ corresponds to $\lambda_{P2}$ and the GameACP or GameAPTC process term $r$ corresponds to $\lambda_{P3}$, then the process term $s \sqcap t \sqcap r$ exactly defines an execution of $\lambda_{P1}$, $\lambda_{P2}$ and $\lambda_{P3}$.

For the strategy $\lambda_{P1}$ corresponding to the GameACP or GameAPTC process term $start\cdot shopping\cdot sPlane \cdot oPlane \cdot (pOnLine \ddagger pOffLine \cdot(ByCheck \ddagger ByBank))$ in Fig.\ref{PurchasingPlayer1}, the strategy $\lambda_{P2}$ corresponding to the GameACP or GameAPTC process term $start\cdot shopping\cdot (sTruck\cdot oTruck\cdot pOnLine \ddagger sTrain\cdot oTrain\cdot pOnLine \ddagger sPlane \cdot oPlane \cdot pOffLine \cdot (ByCheck \ddagger ByBank))$ in Fig.\ref{PurchasingPlayer2}, and the strategy $\lambda_{P3}$ corresponding to the GameACP or GameAPTC process term $start\cdot shopping\cdot (sTruck\cdot oTruck\cdot pOnLine \ddagger sTrain\cdot oTrain\cdot pOnLine \ddagger sPlane \cdot oPlane \cdot (pOnLine \ddagger pOffLine \cdot ByBank))$ in Fig.\ref{PurchasingPlayer3}, the maximal element $start\cdot shopping\cdot sPlane \cdot oPlane \cdot pOffLine \cdot ByBank$ of $\lambda_P \cap \lambda_O$ defines an execution of the process as illustrated in Fig.\ref{Execution2}.

\begin{eqnarray}
&&(start\cdot shopping\cdot sPlane \cdot oPlane \cdot (pOnLine \ddagger pOffLine \cdot(ByCheck \ddagger ByBank)))\nonumber\\
&&\sqcap (start\cdot shopping\cdot (sTruck\cdot oTruck\cdot pOnLine \ddagger sTrain\cdot oTrain\cdot pOnLine \ddagger\nonumber\\
&& sPlane \cdot oPlane \cdot pOffLine \cdot (ByCheck \ddagger ByBank)))\nonumber\\
&&\sqcap (start\cdot shopping\cdot (sTruck\cdot oTruck\cdot pOnLine \ddagger sTrain\cdot oTrain\cdot pOnLine \ddagger\nonumber\\
&& sPlane \cdot oPlane \cdot (pOnLine \ddagger pOffLine \cdot ByBank)))\nonumber\\
&\overset{\text{PO1-PO14}}{=}&start\cdot shopping\cdot sPlane \cdot oPlane \cdot pOffLine \cdot ByBank\nonumber
\end{eqnarray}

\section{Conclusions}

In order to describe game theory and external choice in ACP, we do extensions of ACP and APTC with an opponent's alternative composition operator $\ddagger$ which are called GameACP and GameAPTC. To model the playing process of games, an extension of GameACP or GameAPTC with a playing operator $\sqcap$ and a deadlock constant $\delta$ is also made. And two sound and complete axiomatic system are designed. As a result of axiomatization, GameACP or GameAPTC has several advantages, for example, it has both a proof theory and also a semantics model, it is without any assumptions and restrictions and it can be used in multi-person games naturally. GameACP or GameAPTC can be used to reason about the behaviors of parallel and distributed systems with game theory supported. And also, GameACP or GameAPTC gives games an axiomatization interpretation naturally, this will help people to capture the nature of games.

It must be explained that any computable process can be represented by a process term in ACP (exactly $\textrm{ACP}_\tau$ with guarded linear recursion) \cite{Consistency}. That is, ACP may have the same
expressive power as Turing machine. Although GameACP or GameAPTC can not improve the expressive power of ACP, it still provides an elegant and convenient way to model game theory in ACP.

%\newpage

\label{lastpage}

\end{document}